\documentstyle[12pt,epsf]{article}
\newlength{\dinwidth}
\newlength{\dinmargin}
\newlength{\extraspace}
\newlength{\extraspaces}
\newcommand{\be}{\begin{equation}
\addtolength{\abovedisplayskip}{\extraspaces}
\addtolength{\belowdisplayskip}{\extraspaces}
\addtolength{\abovedisplayshortskip}{\extraspace}
\addtolength{\belowdisplayshortskip}{\extraspace}}
\newcommand{\ee}{\end{equation}}
\newcommand{\bdm}{\begin{displaymath}
\addtolength{\abovedisplayskip}{\extraspaces}
\addtolength{\belowdisplayskip}{\extraspaces}
\addtolength{\abovedisplayshortskip}{\extraspace}
\addtolength{\belowdisplayshortskip}{\extraspace}}
\newcommand{\edm}{\end{displaymath}}
\renewcommand{\thefootnote}{\fnsymbol{footnote}}
\def\simlt{\mathrel{\lower2.5pt\vbox{\lineskip=0pt\baselineskip=0pt
           \hbox{$<$}\hbox{$\sim$}}}}
\def\simgt{\mathrel{\lower2.5pt\vbox{\lineskip=0pt\baselineskip=0pt
           \hbox{$>$}\hbox{$\sim$}}}}
\catcode`@=11
\newcount\@tempcntc
\def\@citex[#1]#2{\if@filesw\immediate\write\@auxout{\string\citation{#2}}\fi
  \@tempcnta\z@\@tempcntb\m@ne\def\@citea{}\@cite{\@for\@citeb:=#2\do
    {\@ifundefined
       {b@\@citeb}{\@citeo\@tempcntb\m@ne\@citea\def\@citea{,}{\bf ?}\@warning
       {Citation `\@citeb' on page \thepage \space undefined}}%
    {\setbox\z@\hbox{\global\@tempcntc0\csname b@\@citeb\endcsname\relax}%
     \ifnum\@tempcntc=\z@ \@citeo\@tempcntb\m@ne
       \@citea\def\@citea{,}\hbox{\csname b@\@citeb\endcsname}%
     \else
      \advance\@tempcntb\@ne
      \ifnum\@tempcntb=\@tempcntc
      \else\advance\@tempcntb\m@ne\@citeo
      \@tempcnta\@tempcntc\@tempcntb\@tempcntc\fi\fi}}\@citeo}{#1}}
\def\@citeo{\ifnum\@tempcnta>\@tempcntb\else\@citea\def\@citea{,}%
  \ifnum\@tempcnta=\@tempcntb\the\@tempcnta\else
   {\advance\@tempcnta\@ne\ifnum\@tempcnta=\@tempcntb \else \def\@citea{--}\fi
    \advance\@tempcnta\m@ne\the\@tempcnta\@citea\the\@tempcntb}\fi\fi}
\catcode`@=12
\newcommand{\SM}{Standard Model}

\newcommand{\br}{branching ratio}

\newcommand{\cL}{{\cal L}}

\newcommand{\pr}{Phys.\ Rev.\ }

\newcommand{\prl}{Phys.\ Rev.\ Lett.\ }
\newcommand{\np}{Nucl.\ Phys.\ {\bf B}}
\newcommand{\pl}{Phys.\ Lett.\ {\bf B}}

\begin{document}
\begin{titlepage}
\begin{flushright}
TUM-HEP-248/96\\
hep-ph/9605272\\
\today \\
\end{flushright}
\vspace{24mm}
\begin{center}
\Large{{\bf Topcolor: A dynamical approach to top-quark mass
generation}}\footnote{Invited 
talk presented at the XXXI Rencontres de Moriond on QCD and High
Energy Hadronic Interactions, Les Arcs, France, 23-30 March 1996.}

\end{center}
\vspace{5mm}
\begin{center}
%\large
Dimitris Kominis\footnote{e-mail address: kominis@physik.tu-muenchen.de}
\\*[3.5mm]
{\normalsize\it Institut f\"ur Theoretische Physik, 
Technische Universit\"at M\"unchen,}\\
{\normalsize\it James-Franck-Stra\ss e, 85748 Garching, Germany }
\end{center}
\vspace{2cm}
\thispagestyle{empty}
\begin{abstract}
In Topcolor theories the mass of the top quark is generated dynamically by new
strong interactions coupling to 
the third generation fermions. This leads to
potentially observable effects in flavor-changing neutral current processes
involving heavy flavored mesons,
as well as in the production of heavy quarks at high-energy colliders. 
Recent theoretical
developments and potential phenomenological constraints on these models are
also briefly reviewed.

\end{abstract}
\end{titlepage}
\newpage

\renewcommand{\thefootnote}{\arabic{footnote}}
\setcounter{footnote}{0}
\setcounter{page}{2}

\section{Introduction} 
\vspace{-0.2cm}

The present data confirm the existence of a top quark with a mass of
approximately 170~GeV and with properties consistent with the predictions of
QCD \cite{tev}. Naturally, this raises the 
question ``Why is the top quark so heavy?". This talk deals with one
approach to this problem, usually referred to as Topcolor (TopC). Our
discussion will be mostly confined to one particular model; however,
this will be regarded as a prototype of theories in which
the large top quark mass has a dynamical origin. In such theories, the top
quark participates in a new strong interaction which causes it to
condense and
thereby acquire a large mass, in analogy to the {\it constituent} mass of the
light quarks in QCD. The talk consists of an outline of the simplest TopC model,
a brief review of recent theoretical work and finally a
discussion of phenomenology, divided into two broad categories. 
The first deals with flavor-changing neutral currents (FCNC's) which arise
when the third generation is treated preferentially. (Note that the $b$-quark,
or at least its left-handed component, is necessarily dragged into 
the new dynamics because it is the $SU(2)_L$ partner of top.) No natural GIM
mechanism can be invoked to suppress the FCNC's in this case
because the gauge interactions are no
longer flavor-symmetric. The second leg of the phenomenology discussion
concerns the production of heavy flavors at high-energy colliders. The
consequences of TopC on some
observables on the $Z$-peak are also commented upon.

\vspace{-3.0mm}
\section{A Topcolor Model}
\vspace{-2.2mm}
The simplest TopC model is due to Hill \cite{hill}:
At scales higher than a few
TeV the gauge symmetry group is extended to $SU(3)_1 \times SU(3)_2 \times
U(1)_1 \times U(1)_2 \times SU(2)_L$. The groups $SU(3)_1$ and $U(1)_1$ are
stronger and couple to the third generation only, while 
$SU(3)_2$ and $U(1)_2$ couple to the first two generations. At a scale of $\sim
1$~TeV the $SU(3)_1 \times SU(3)_2$  breaks down spontaneously to its diagonal
$SU(3)$ subgroup, identified with QCD, giving rise
%through the Higgs mechanism 
to a color octet of massive gauge bosons $B^A$, called
colorons. Their mass is determined by the scale of symmetry breaking and the
$SU(3)$ couplings and is therefore of the order of 1~TeV. The colorons
couple to quarks in the following way:
\be
\cL _{B} = g_3 B_{\mu}^A \left(-\tan \theta \sum_{q \ne t,b}
\bar{q}\gamma^{\mu}\frac{\lambda ^A}{2}q + \cot \theta \sum_{Q=t,b} \bar{Q} 
\gamma^{\mu}\frac{\lambda ^A}{2} Q\right)
\label{Bqq}
\ee
where $g_3$ is the QCD coupling, $\lambda^A$ are the Gell-Mann 
matrices and $\theta$ is a small
mixing angle given by $\tan\theta = h_2/h_1 \ll 1$, where $h_i$ is the
$SU(3)_i$ gauge
coupling. Similarly the $U(1)_1 \times
U(1)_2$ is assumed to break down to the $U(1)$ of hypercharge producing 
a massive $Z^\prime$ boson with strong couplings to
the third family quarks {\it and} leptons:
%\footnote{This is necessary in order to
%cancel gauge anomalies without introducing new fermions.}. 
%but weakly to the first two families:
\be
\cL _{Z^\prime} = g_1 Z^{\prime}_{\mu} \left(-\tan \theta^\prime 
\sum_{f\ne t,b,\tau,\nu_{\tau}} \bar{f}
\gamma^{\mu}\frac{Y_f}{2} f + \cot \theta^\prime \sum_{F=t,b,\tau,\nu_{\tau}}
\bar{F} \gamma^{\mu} \frac{Y_F}{2} F \right)
\label{zff}
\ee
Here $\tan\theta^\prime = q_2/q_1 \ll 1$, $q_i$ being the $U(1)_i$ gauge
coupling, $g_1 $
%= q_1 q_2/\sqrt{q_1^2 + q_2^2}
is the ordinary hypercharge
gauge coupling and $Y_f$ is the hypercharge of (chiral) fermion $f$. The
coloron mediates a strong attractive interaction between $t$ and $\bar{t}$ as
well as between $b$ and $\bar{b}$, while the $Z^\prime$ interaction is
attractive in the $t$ channel but {\em repulsive} in the $b$ channel. 
This is due
to the different hypercharge assignments. The combined effect is then that the
attractive force between $t$ and $\bar{t}$ 
can exceed a critical value for the formation of a $\langle
\bar{t}t \rangle$ condensate (and hence a large dynamical top quark mass),
while the more weakly coupled bottom sector remains subcritical ($\langle
\bar{b}b \rangle = 0, m_b^{\rm dyn} =0 $). 
%[The introduction of the strong $U(1)$ group was therefore intended to provide
%the necessary custodial isospin breaking in the third generation.]

The $\langle \bar{t}t \rangle$ condensate spontaneously breaks the chiral
symmetries associated with the top quark and gives rise to a set of three
Goldstone bosons, henceforth referred to as top-pions. 
In contrast to the early models of top condensation \cite{bhl}, which
attempted to identify them with the
Goldstone bosons of electroweak symmetry breaking, these top-pions are new,
physical degrees of freedom\footnote{This is a consequence of the low scale
$\Lambda$ of
new interactions, which ensures a natural hierarchy $m_t\!:\!\Lambda$, 
but forces the top-pion decay
constant to a value too low to account for the whole of electroweak
symmetry breaking.}.
Actually the top-pions are not exactly massless, because
the top quark mass is allowed to have a small {\it explicit}
component ($\sim 1$~GeV) generated by the same
mechanism which is responsible for the other fermion masses. Simple
estimates \cite{hill} indicate that this leads to a top-pion mass of the
order of 200~GeV. 

In this model the issues of
TopC and electroweak symmetry breaking are not addressed.
Recent attempts to construct theories where these symmetries are dynamically
broken by a technicolor sector are outlined in the next section.

\vspace{-3.0mm}
\section{Topcolor-Assisted Technicolor}
\vspace{-2.2mm}

In technicolor \cite{tc} the electroweak symmetry is broken by 
condensates $\langle \bar{T}_L T_R \rangle$ of a new kind of fermions
(``technifermions"), 
which interact via the technicolor force.
The symmetry breaking is then communicated to the
ordinary quarks and leptons by interactions which couple them to technifermions
at a scale of the order of 100~TeV. This mechanism \cite{etc}, 
known as extended technicolor
(ETC), generates the quark and lepton masses in this scheme. However, within
the context of traditional ETC, it has been very difficult to construct a
technifermion sector which, on one hand, contains sufficient custodial
isospin breaking to explain the large $t-b$ mass difference, while on the other
prevents this isospin breaking from diffusing into the electroweak sector
and upsetting the value of the $\rho$ parameter \cite{rho}. 
Now, by assigning the
task of top quark mass generation to TopC, this problem may be to a
considerable extent alleviated.
Models which implement this idea have been constructed in
Ref. 7. They feature the following basic elements:

$\bullet$ Technifermion condensates are responsible 
for the breaking of not only
the electroweak symmetries but also the TopC gauge symmetries. \\
%so no extra dynamics needs to be introduced for this purpose.
\indent
$\bullet$  In fact, some condensates break 
the electroweak and TopC symmetries {\it
simultaneously}. This turns out to be crucial to the
generation of mixings of the correct size between quarks of the third and first
two families.\\ \indent
$\bullet$  All technifermions couple to the strong TopC gauge groups 
$SU(3)_1$ and $U(1)_1$ in an
isospin symmetric fashion. In this way large corrections to the electroweak
$\rho$ parameter are avoided \cite{cdt}.\\ \indent
$\bullet$  The models are ``complete" in the sense that the specified fermion
and technifermion content is sufficient to cancel all gauge
anomalies. 
The ETC gauge group, however, is not specified; the ETC interactions 
are simulated by ``phenomenological" four-fermion terms consistent with all
gauge symmetries.
\\ \indent
$\bullet$  The pseudo-Goldstone bosons which typically arise in ETC models from
the spontaneous breakdown of large chiral symmetry groups all acquire here a
``safe" mass of the order of 250~GeV or larger. \\ \indent
$\bullet$  In a departure from the basic postulates of the simplest TopC
model of the previous section, both heavy and light fermions couple
to the stronger TopC $U(1)_1$ gauge group.
A detailed analysis of the manifold consequences that this may have (e.g. on 
atomic parity violation observables, 
the leptonic width of the $Z$ boson or jet
production at the Tevatron, to name but a few) has not been yet
carried out. Therefore, for the phenomenological discussion 
that follows we shall
return to the ``minimal" TopC scheme of the previous section.

\vspace{-3.0mm}
\section{Phenomenology}
\vspace{-1.8mm}
\subsection{Low-Energy FCNC Phenomenology}
\vspace{-1.2mm}

In the absence of a natural GIM mechanism, FCNC processes
mediated by a coloron or a $Z^\prime$ occur at tree level. Since
these massive gauge bosons couple strongly to the third
family quarks, 
they can have an appreciable effect
on rare $B$-meson decays, where the \SM\ contribution arises only at
loop level and is accordingly suppressed. 
These processes
depend on the mixing angles that effect the transition from weak to mass
eigenstates. There are in general four such mixing matrices, $U_L,U_R,
D_L,D_R$, 
corresponding to up-type, down-type, left- or right-handed quark field
rotations. In contrast to the Standard Model, where only the combination
$V\equiv U_L^\dagger D_L$ which defines the Cabibbo-Kobayashi-Maskawa (CKM)
matrix is meaningful, here all four mixing matrices are well-defined
entities. Their elements cannot all be deduced from the CKM matrix and are
therefore unknown parameters of the theory. For the purposes of making
estimates, we assume that these mixing matrices are roughly equal:
\be
U_L \sim D_L \sim U_R \sim D_R \sim V^{1/2}
\label{rootkm}
\ee
We will further assume typical values for the mass and couplings of the
$Z^\prime$: $M_{Z^\prime}\simeq 1$~TeV, $\kappa_1 \equiv g_1^2
\cot^2\theta^\prime /4\pi \approx 1$ (see eq. (\ref{zff})).
Note that this coupling is
non-perturbative and thus the numbers obtained are at best order-of-magnitude
estimates. Table~1 lists some observables that are particularly sensitive to
the TopC dynamics \cite{bbhk} and compares the TopC and Standard Model
predictions. The following additional remarks are in order: 

\begin{table}
\begin{center}
\begin{tabular}{|l|c|c|}     \hline
Process & TopC BR & SM BR \\ \hline
$B_s^0 \rightarrow \tau^+\tau^-$ & $\sim 2\times 10^{-5}$ & $\sim 10^{-6}$ \\
\hline
$B \rightarrow X_s \tau^+ \tau^-$ & $0.9\times 10^{-4}$ & $3.7\times 10^{-7}$
\\ \hline
$B \rightarrow X_s \mu^+ \mu^-$ & $6.0\times 10^{-6}$ & $6.3\times 10^{-6}$ \\
\hline
$B \rightarrow X_s \nu \bar{\nu}$ & $2.7\times 10^{-4}$ & $4.5\times 10^{-5}$
\\ \hline
$K^+ \rightarrow \pi^+ \nu \bar{\nu}$ & $\sim 2.5\times 10^{-10}$&$\sim
10^{-10}$ \\ \hline
\end{tabular}
\caption{Rare decay processes and their branching ratios in TopC and 
in the \SM.}
%(see text). $X_s$ stands for an inclusive strange state.}

\end{center}
\end{table}
\vspace{-1mm}
(i) There are no published data on the processes $B_s^0 \rightarrow \tau^+
    \tau^-$ and $B \rightarrow X_s \tau^+ \tau^-$, but the results on Table~1
    show that an experimental effort in this direction is very
    desirable. 
%   (For different $Z^\prime$ masses and couplings, the TopC
%    prediction scales like $\sim \kappa_1^2 /
%    M_{Z^\prime}^4$.)
(ii) In $B \rightarrow X_s \mu^+ \mu^-$ the deviation from the \SM\
     prediction is rather small, because the $Z^\prime$ which mediates the
     TopC contribution couples only weakly to muons (see eq. (\ref{zff})). 
     Even so, the
     angular distribution of the final state leptons provides
     a signal quite distinct from the \SM\ \cite{gb}. 
     This is a consequence of
     the fact that in TopC this decay can proceed via effective operators of a
     chirality structure which is forbidden in the \SM. Figure~1
     illustrates this point in the case of the {\it exclusive} decay
     $B \rightarrow K^* \mu^+ \mu^-$. On this graph, the forward-backward
     asymmetry $A_{FB}$ of the muons, defined relative to the direction of
     motion of the $B$ meson in the $\mu^+ \mu^-$ c.m. frame, is plotted as a
     function of the dimuon invariant mass $m_{ll}$. 
\begin{figure}
\vspace{0.5cm}
%\special{psfile=Afb_SM.eps
%          angle=0 hscale=40 vscale=50 hoffset=-30 voffset=-210}
\includegraphics{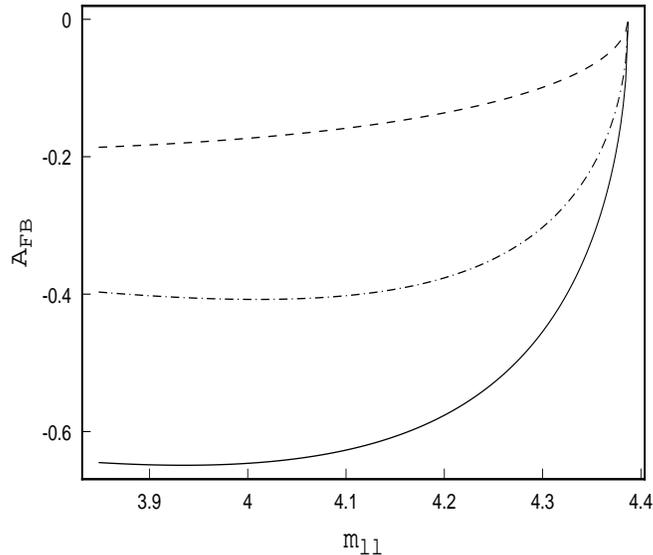}
\vspace{0.0cm}\hspace{3cm}\hspace{8.5cm}\vspace{1cm}
\vspace{6cm}
\small
\caption[]{Forward-backward asymmetry of the muons in $B \rightarrow K^* \mu^+
\mu^-$ as a function of the dimuon invariant mass. 
Solid: \SM; dot-dash: TopC with $M_{Z^\prime}=$1~TeV; dashes: TopC with
$M_{Z^\prime}=$0.5~TeV. From Ref. 10.}
\normalsize 
\label{f_as1}
\end{figure}
     Even for a 1~TeV $Z^\prime$ boson
     the deviation from the \SM\ is striking.
%     even though the total rate for
%     this decay is not significantly different in the two models.
     It is possible 
     that with the Main Injector, and
     assuming \SM\ branching ratios, CDF will be able to collect enough events
     to make an angular analysis of this kind possible. 
(iii) The decay $B \rightarrow X_s \nu \bar{\nu}$ receives a significant
contribution from TopC when $\nu = \nu_{\tau}$. Here again
there are no direct experimental limits on the \br, but a preliminary upper
bound of $3.9\times 10^{-4}$ has appeared in the
literature \cite{grossmann}.   
(iv) Finally, the process $K^+ \rightarrow \pi^+ \nu \bar{\nu}$ can also be of
interest, given that experiments which can reach \SM\ sensitivity are planned
for the near future \cite{kpinn}. 
Here, however, the new physics effects are not pronounced
because only quarks of lower generations are involved. 

All of the decays displayed in Table~1 are mediated by a $Z^\prime$ boson and
are accordingly suppressed by powers of its mass. In TopC, however, FCNC
processes are sometimes amplified by the effects of tightly bound states which
are light compared to the scale of new interactions. 
It is estimated \cite{me} that scalar bound states coupling
predominantly to the $b$-quark may exist in the few hundred~GeV mass
range. Under the assumptions of eq. (\ref{rootkm}), exchange of
these particles would lead to a 
violation of the observed mass difference in the neutral $B_d$
meson system by approximately two orders of magnitude! Assuming the estimates
are reliable, this constraint can only be evaded if at least one of the mixing
factors $D_L^{bd}$ and $D_R^{bd}$ is suppressed. (This does not affect any of
the processes in Table~1.) It is interesting to note that such a suppression
of $D_R^{bd}$ occurs 
naturally in the context of the Topcolor-Technicolor model
of Section~3. 
The top-pions mediate in an analogous fashion $D$ to $\bar{D}$
transitions. Here the TopC prediction, under the assumption (\ref{rootkm}),
is $\Delta m_D \approx 2\times 10^{-14}$~GeV \cite{bbhk}, 
still below the experimental
limit of $1.3\times 10^{-13}$~GeV, but within reach in future high statistics
experiments.%\cite{charm}.

\vspace{-1.2mm}
\subsection{Heavy Flavor Production}
\vspace{-1.2mm}

In TopC pair production of third generation quarks can proceed via a coloron 
in the $s$-channel \cite{hillp,hz}. ($Z^\prime$ exchange is subdominant.)
At the Tevatron, $t\bar{t}$ production requires the scattering of partons of
large $x$ and hence $q\bar{q}$ annihilation is the dominant production
mechanism (with gluon fusion contributing only about 10\% of the total
rate). The situation is reversed at the LHC, where the 
center-of-mass energy is
much higher and partons of lower $x$ play the dominant role. Since the coloron
can only be produced by quark-antiquark annihilation,
it follows that its
effects will be more significant at the Tevatron. This is illustrated in
Fig.~2, where the ratio of the total $t\bar{t}$ production cross-section
in TopC to the same cross-section in the \SM\ is plotted as a function of the
coloron mass. 
Results are shown for both the Tevatron and the LHC. 
A top quark mass of 175~GeV is assumed. As the error bars on the top 
quark mass
and production cross-section from the CDF and D0 shrink, limits on the coloron
mass will arise. 
\begin{figure}
%[htbf]
\epsfxsize 2.0in
\vspace{0.25in}
\centerline{\epsffile{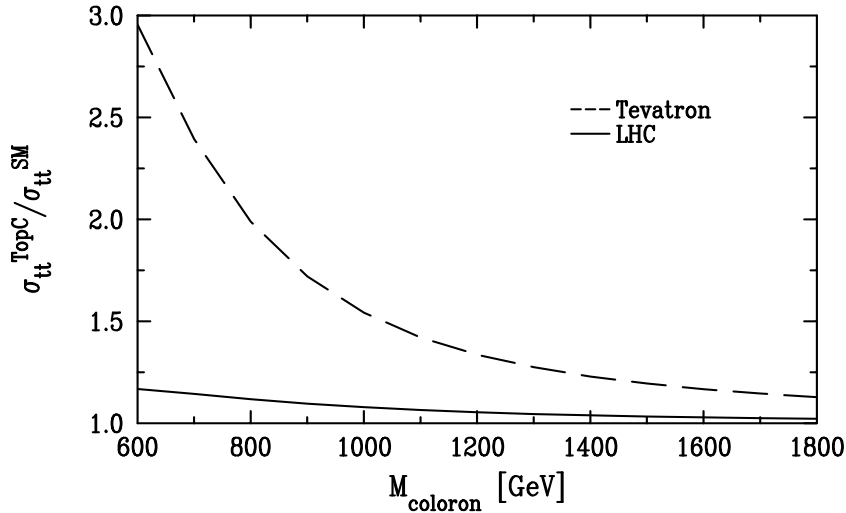}}
\caption{Comparison of top-pair production cross-section in TopC and 
in the \SM\ at the Tevatron and the LHC. (Inspired by Ref. 15.)}
\label{gg}
\end{figure}

The TopC gauge bosons will also cause distortions to the shape of various 
distributions \cite{hillp}. 
The effect will be particularly noticeable in the $t\bar{t}$ invariant mass 
distribution, where the coloron
will appear as a broad shoulder over the continuum. The same remarks
apply to $b\bar{b}$ production. 
%because the coloron
%couples to $b\bar{b}$ with the same strength as it does to $t\bar{t}$. 
An experimental search in this channel has set a first bound of 
$M_B \simgt 370$~GeV 
%at 95\% C.L.
on the coloron of a slightly modified version of
TopC \cite{cdfcoloron}.
%(This bound is based on the assumption that the
%width of the coloron is equal to half its mass.)

The CDF collaboration has recently reported an excess in
the inclusive jet cross-section at high $p_T$ relative to \SM\
expectations \cite{cdfjets}. The conventional TopC model
of Section~2 can only lead to an excess of $b$-jets, because the new heavy
gauge bosons decay predominantly to third generation quarks.
However, the observed effect could be attributed to the massive gauge bosons of
alternative TopC models \cite{topc2}, in which
some or all of the light quarks
participate in the new strong interactions as well. Detailed fits of
TopC models to the CDF data have not, however, been performed so far.

Finally, it should be mentioned that the new TopC interactions may induce
potentially large corrections to $Z$-pole observables. Lepton universality, for
instance, is violated because the $Z^\prime$, which mixes with the $Z$, couples
more strongly to the $\tau$ lepton. Hence a lower bound of approximately 1~TeV
%at $3\sigma$ 
on the $Z^\prime$ mass can be derived if $\kappa_1=1$. This bound
scales roughly with the square-root of $\kappa_1$. 
Regarding $R_b \equiv \Gamma
(Z\rightarrow b\bar{b}) / \Gamma (Z\rightarrow hadrons)$, 
the positive coloron vertex correction \cite{hz} may be offset by negative
contributions due to top-pions \cite{inprog}. 
The issue is still under study.

%\vspace{-3.0mm}
%\section{Conclusions}
%\vspace{-2.2mm}
In this talk, I reviewed recent 
developments in theories in which the large top
quark mass is tied to new strong dynamics of the third generation. The aim was
more to explore the sensitivity of various physical processes to this kind of
new physics than to promote any particular realization of it. It was shown that
this new dynamics can manifest itself
%in some cases rather dramatically, 
in FCNC processes, such as $B-\bar{B}$ or $D-\bar{D}$
mixing and leptonic and semileptonic $B$ meson decays, as well as in the
production of heavy flavors at high-energy colliders. 

I thank the organizers of the XXXIst Rencontres de Moriond for their kind
invitation. I am indebted to K. Lane for valuable discussions and, naturally,
to my 
collaborators C. Hill, G. Burdman and G. Buchalla. This work was supported in
part by the German DFG under contract number Li519/2-1.

\end{document}